\documentclass[12pt, amsmath,amssymb,aps]{revtex4}
\usepackage{graphicx,subeqnarray,dcolumn,bm,amsmath,epstopdf}
\usepackage{amssymb,commath,eqnarray,gensymb,physics}
\usepackage{color,siunitx}
\usepackage[colorlinks=true, citecolor=red, linkcolor=blue,urlcolor=blue ]{hyperref}

\DeclareMathAlphabet{\mathsfbi}{OT1}{\sfdefault}{bx}{n}

\renewcommand{\vec}[1]{\mathbf{#1}}

\begin{document}
\title{Helical micropumps near surfaces}
\author{Justas Dauparas}
\author{Debasish Das}
\author{Eric Lauga}
\email{e.lauga@damtp.cam.ac.uk}
\affiliation{Department of Applied Mathematics and Theoretical Physics, University of Cambridge, CB3 0WA, United Kingdom}
\date{\today}

\begin{abstract}
Recent experiments proposed to use confined bacteria in order to generate flows near surfaces. We develop a mathematical and a computational model of this fluid transport using a linear superposition of fundamental flow singularities. The rotation of a helical  bacterial flagellum induces both a force and a torque on the surrounding fluid, both of which lead to a net flow along the surface. The combined flow is in general directed at an angle to the axis of the flagellar filament. The optimal pumping is thus achieved when bacteria are tilted with respect to the  direction in which one wants to move the fluid, in good agreement with experimental results. We further investigate the optimal helical shapes to be used as micropumps near surfaces and show that  bacterial flagella are nearly optimal, a result which  could be relevant to the expansion of bacterial swarms.
\end{abstract}   
\maketitle
 \section{Introduction}

Transporting fluid on small scales is an important but complex problem. Various pumping mechanisms are observed in nature, from   fluid being pushed by muscles in the walls of the lymphatic vessels~\cite{kunert2015mechanobiological},   sodium-potassium pumps in the membranes of cells~\cite{glynn1993annual} and efficient osmotic pumping in plants~\cite{jensen2012universality} to cilia-induced flows in major organs including the brain~\cite{faubel2016cilia}, the respiratory system~\cite{fahy2010airway} and   reproductive tracts~\cite{spassky2017development}. 

Some of these transport mechanisms inspired the design of pumps for microfuidic systems which may be used for biological and chemical sensing, drug delivery, molecular separation, amplification, and sequencing~\cite{nisar2008mems}. There are two major classes of such pumps: mechanical displacement pumps, which apply forces to fluids via moving boundaries, and electro- and magnetokinetic pumps which provide energy to fluids continuously and as a result generate flow~\cite{abhari2012comprehensive}. Mechanical or electrokinetic micropumps are used in the majority of lab-on-a-chip devices that require powered fluid flow, but the technical challenges and the requirement
of external power associated with these pumping devices have impeded their miniaturisation. 

Self-powered micropumps
have been   designed to   address some of these issues~\cite{zhou2016chemistry}. One of the solutions is to build bio-hybrid cell-based actuators. Recently, there has been a growing interest in using live biological cells to produce devices that work as microscopic gears~\cite{sokolov2010swimming}, microrotors~\cite{di2012hydrodynamic}, micropumps~\cite{kim2008microfluidic}, microswimmers~\cite{steager2011electrokinetic}, and microwalkers~\cite{carlsen2014bio}. Bacterial carpets, which are surface arrays of fixed bacteria~\cite{uchida2010synchronization, martindale2017autonomously},  can also create linear and rotational flows~\cite{darnton2004moving} and enhance mixing~\cite{kim2007use}. 

In this work, we focus on bacterial micropumps, with their potential to be used for  automation of chemistry and biology~\cite{squires2005microfluidics}. In the one of the first bacterial micropump experiments, \textit{Escherichia~coli} (\textit{E.~coli}) cells were tethered to microchannel walls by a single flagellar filament and the bodies of the cells would rotate at about 10~rps, thereby pumping fluid from one end of the channel to the other~\cite{tung2003cellular, tung2006microscale}. In another experiment, cells were attached to the surface by their
bodies with flagella free to rotate~\cite{kim2008microfluidic},  demonstrating that the bacteria are able to self-organise, generating a collective flow that can pump fluid autonomously through a microfabricated channel. Furthermore, the addition
of glucose to the working buffer raises the metabolic activity of the bacterial carpet, resulting in enhanced pumping performance.
 
In this paper we consider the bacterial pumping system described in a recent experimental paper  where  \textit{E.~coli}  bacteria were confined within micro-fabricated structures in a prescribed geometrical configuration  and whose flagellar rotation collectively generated flow that can transport materials along designed trajectories~\cite{gao2015using}. This study naturally raises the question of (i)  how the performance of the resulting pumps depends on the detailed geometrical characteristics of the microscopic cages in which the cells are trapped and (ii) how to optimise them. In order to answer these questions, we use the mathematical techniques of resistive-force theory~\cite{lighthill1976flagellar} and  slender-body theory~\cite{johnson1980, tornberg2004simulating} to model the flow induced by the rotating flagellar filaments of the trapped bacteria. We first quantify how the magnitude and direction of the bacteria-driven  pump   depends on the configuration of the flagella and then address the geometrical optimisation of the helical shapes used to generate the flow.

\section{Flux induced by helix rotating above a wall}
In this section we derive a mathematical model  to compute the flux produced by a fixed bacterium whose flagellar filaments are rotating above a no-slip wall. Both resistive-force theory (analytical approach) and  slender-body theory (computational approach) are used   and we compare our results with experimental measurements of Ref.~\cite{gao2015using}.

\subsection{Flux due to singularities above a wall}
\label{IIA}

\begin{figure}[t!]
	\includegraphics[width=0.45\textwidth]{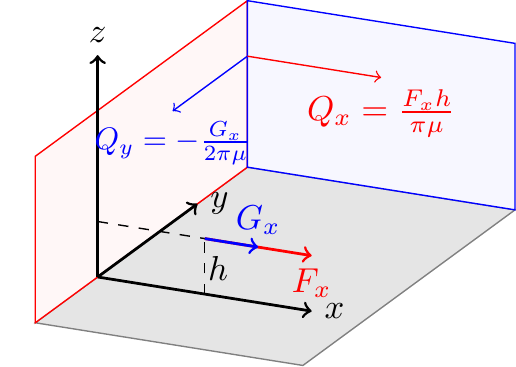}
	\caption {A point force $\vec{F}=(F_x,0,0)$ and a point torque $\vec{G}=(G_x,0,0)$ located at $(0,0,h)$ above the infinite plane at $z=0$ generate a net hydrodynamic flux  $\vec{Q}=F_xh/(\pi \mu) \vec{e}_x-G_x/(2\pi\mu)\vec{e}_y$.}
	\label{fig:19}
\end{figure}

Consider a Cartesian coordinate system $(x,y,z)$ with unit vectors $(\vec{e}_x,\vec{e}_y,\vec{e}_z)$, as illustrated in Fig.~\ref{fig:19}. 
A point force (Stokeslet) $\vec{F}=F_x \vec{e}_x$ is located at a height $h$ above an infinite no-slip wall with  normal $\vec{e}_z$ in a fluid with a dynamic viscosity $\mu$. Assuming the flow to have no inertia, which is a reasonable assumption on the small length scales of bacteria~\cite{gao2015using}, the resulting solution to the incompressible Stokes equations has a velocity field denoted by $\vec{u}$. It is a classical (and exact) result that this point force will then produce a net flux (i.e.~flow rate)~\cite{blake1974,blake1974fundamental}  given by 
\begin{align}\label{QF}
\vec{Q}_{F}=&\int_0^\infty\int_{-\infty}^\infty (\vec{u}\cdot \vec{e}_x)\vec{e}_x dydz+\int_0^\infty\int_{-\infty}^\infty (\vec{u}\cdot \vec{e}_y)\vec{e}_y dxdz\\
\nonumber
&+\int_{-\infty}^\infty\int_{-\infty}^\infty (\vec{u}\cdot \vec{e}_z)\vec{e}_z dxdy\\
\nonumber
=&\frac{F_{x}h}{\pi \mu}\vec{e}_x,
\end{align}
which is  in the same direction as the force (see Fig.~\ref{fig:19}). The calculation for the flux is summarised in Appendix~\ref{appendix:force}. Physically, while the flow induced by a point force decays as $1/r$, a point force near a wall decays faster as $1/r^2$ and since the area of integration scales as distance squared, a net flow rate is obtained in Eq.~\eqref{QF}.

Similarly, a point torque (Rotlet) $\vec{G}=G_x \vec{e}_x$, placed at the same location will produce exactly the net flux
\begin{align}
\vec{Q}_G=-\frac{G_x}{2\pi \mu}\vec{e}_y,
\end{align}
which points  in the direction perpendicular to both the torque and the surface (see the derivation in Appendix~\ref{appendix:torque}). 
\subsection{Flagellar flows as one singularity}
Most motile bacteria, including \textit{E.~coli}, are propelled by the rotation  of  helical flagellar filaments~\cite{berg2008coli}. Each filament is attached to the cell body via a short flexible hook itself  connected to a bacterial rotary motor. The so-called `normal' shape of the flagellar filament is a left-handed helix, typically  rotated in the counter-clockwise direction when looked from behind the cell. An \textit{E.~coli} bacterium has approximately 4 flagella on the body~\cite{turner2000real} which normally  bundle in the back of the cell forming  a thick effective helix~\cite{berg2008coli}.
\begin{figure}[t]
\includegraphics[width=0.55\textwidth]{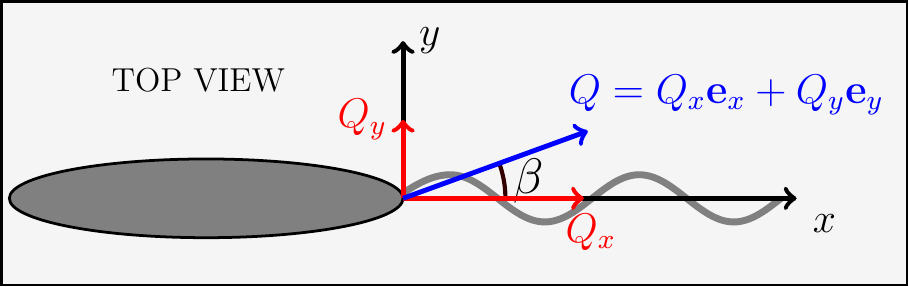}
\includegraphics[width=0.55\textwidth]{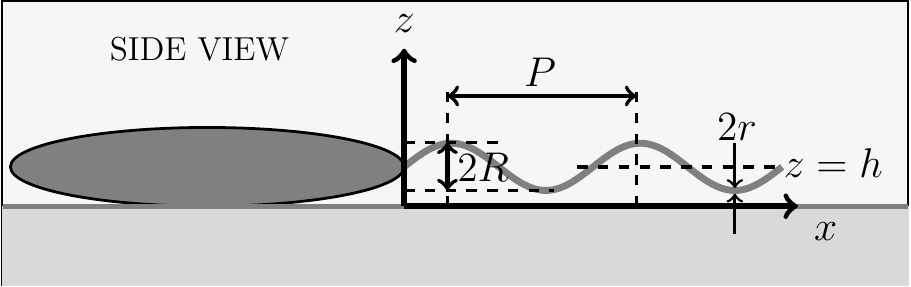}
\caption {Top and side view of a bacterium with a helical bundle of flagellar filaments. The helical bundle has  pitch $P$,   radius $R$,  while the bundle radius is denoted by $r$. 
The axis of the  helix  is positioned at height $h$ above the surface. With the cell body stuck,  the rotating bundle  creates a net flow  parallel to the surface with flow rate  $\vec{Q}=Q_x \vec{e}_x+Q_y\vec{e}_y$, and thus at an angle $\tan{\beta}=Q_y/Q_x$   to the $x$ direction.}
\label{fig:1}
\end{figure}
The simplest model for the flow induced by a rotating bundle of flagellar filaments near a surface consists of computing the net force,  $\vec{F}$, and torque,  $\vec{G}$ which the bundle applies on the fluid and use the result for from the previous section. In this simple model, we neglect the presence of the cell body on the fluid flow because only a small portion of the flagellar bundle is near the cell body while the majority  of the bundle is closer to the no-slip wall. Let us then assume we have a perfect helix   parallel to a no-slip wall whose axis is at height $z=h$ with  the wall   at $z=0$ (see Fig.~\ref{fig:1}). 

In order to relate the values of $\vec{F}$ and   $\vec{G}$ to the geometry of the helical bundle, we use resistive-force
theory (RFT) which captures the density of hydrodynamic forces acting  on slender filaments in Stokes flows~\cite{lighthill1976flagellar}. We denote  by 
$\xi_{\perp}$  the drag coefficient (i.e.~force per unit length) for translation of a portion of the  filament bundle locally perpendicular to its tangent and 
 $\xi_{\parallel}$  the drag coefficient in the parallel direction and we write their ratio as  $\rho=\xi_{\parallel}/\xi_{\perp}$.  We will use the drag coefficients obtained by Lighthill~\cite{lighthill1976flagellar} as
\begin{align}
 \xi_{\parallel} &=\frac{2\pi\mu}{\ln{(0.18 P\sec{\Psi}/r)}},\quad
 \xi_{\perp} =\frac{4\pi\mu}{1/2+\ln{(0.18 P\sec{\Psi}/r)}},
\end{align}
where we denote $\mu$ the dynamic viscosity of fluid, $r$ the bundle radius,
  $R$ the helix radius (i.e.~the radius of the cylinder on which its centreline is coiled), $P$ its pitch  and  $\Psi$  the pitch angle (i.e.~$\tan\Psi={2\pi R}/{P}$), as shown in Fig.~\ref{fig:1}.
 The axis of the helix axis is assumed to be directed  along the $x$ direction and we use  cylindrical coordinates $(\sigma,\theta,x)$ around the helix axis. 

It is a classical result~\cite{lauga2009hydrodynamics} that the  force per unit length acting on the fluid by a counter-clockwise (CCW, i.e.~in the positive $x$ direction) rotating left-handed helix has components
\begin{align}
f_{\theta}&=\xi_{\perp} R \omega (\cos^2{\Psi}+\rho \sin^2{\Psi)},\\
 f_x&=\xi_{\perp} R \omega (1-\rho) \sin{\Psi} \cos{\Psi},\\
 f_{\sigma}&=0,
\end{align}
where  $\omega$ denotes the angular velocity of the helix. 
The total force acting on the fluid and the total torque about the  axis of the helix is obtained by integrating the force density along the arc length $0\leq s \leq L$ leading to
\begin{align}\label{Fx}
 F_x&=\int_0^{L}f_x ds=\xi_{\perp} R \omega (1-\rho) \sin{\Psi} \cos{\Psi} L,\\
\label{Gx} G_x&=\int_0^{L}f_{\theta}R ds =\xi_{\perp} R^2 \omega (\cos^2{\Psi}+\rho \sin^2{\Psi)} L,
\end{align}
where $L$ is the total contour  length of the helix.
Using the result of \S\ref{IIA}, these force and torque induce a net flow rate of 
\begin{align}\label{eq6}
\vec{Q}&=\frac{F_x h}{\pi\mu} \vec{e}_x-\frac{G_x}{2\pi\mu} \vec{e}_y.
\end{align}
Using $Q$ to denote the magnitude of the flux and $\beta$ the angle between the net flow and the $x$ axis (see Fig.~\ref{fig:1}), Eq.~\eqref{eq6} leads to 
\begin{equation}\label{eq7}
Q^2=\left(\frac{F_x h}{\pi\mu}\right)^2+\left(\frac{G_x}{2\pi\mu}\right)^{2},\quad\beta=-\arctan{\left(\frac{G_x}{2F_x h}\right)}.
\end{equation}
If we substitute the expression for the force and the torque from Eq.~\eqref{Fx}-\eqref{Gx} the model leads to the prediction 
\begin{align}
\label{eq:qm}
Q^2&=\left(\frac{\xi_{\perp} R \omega (1-\rho) \sin{\Psi} \cos{\Psi} L h}{\pi\mu}\right)^2+\left(\frac{\xi_{\perp} R^2 \omega (\cos^2{\Psi}+\rho \sin^2{\Psi)} L}{2\pi\mu}\right)^{2},\\
\beta&=-\arctan{\left(\frac{(\cos^2{\Psi}+\rho \sin^2{\Psi)}R}{ 2 (1-\rho) \sin{\Psi} \cos{\Psi}h }\right)}.
\label{eq:beta}
\end{align}
\subsection{Flagellar flows as a superposition of singularities}
The approach in the previous section modelled the helix as a single force and torque singularity. Alternatively,  we may write the helix as a superposition of flow singularities. The centreline of the left-handed helix may be parametrised as $(x_0(s,t),y_0(s,t),z_0(s,t))$  with
\begin{align}
      x_0(s,t)&= s \cos\Psi,\\
      y_0(s,t)&=-R \sin{\left(\omega t - 2\pi \frac{s}{P} \cos\Psi \right)},\\
      z_0(s,t)&=h-R\cos{\left(\omega t - 2\pi \frac{s}{P} \cos\Psi\right)},
\end{align}
where $s$ denotes the arclength along the helix centreline and $t$ time. 
As a result the total flow rate has   
\begin{align}
Q_x(t)&=\frac{1}{\pi\mu}\int_0^{L} f_x(s) z_0(s,t) ds,\\
Q_y(t)&=\frac{1}{\pi\mu}\int_0^{L} f_y(s,t) z_0(s,t) ds=\frac{1}{\pi\mu}\int_0^{L} f_{\theta}(s)\cos{\left(\omega t-2\pi\frac{s}{P}\cos\Psi\right)} z_0(s) ds.
\end{align}
If we average over time then we get the same results as before considering the point force and the point torque namely Eqs.~\eqref{eq:qm}-\eqref{eq:beta}.
\subsection{Numerical approach: Slender-body theory}
In parallel to the analytical approach we may use the improved, computational model  termed slender-body theory~\cite{johnson1980}. Under this framework, there is a   linear relationship between the rigid body motion defined by the centreline velocity $\vec{u}$ and the hydrodynamic force density $\vec{f}$ on the fluid formally written as~\cite{tornberg2004simulating}
\begin{align}\label{sbt}
8\pi \mu \vec{u}(\vec{x},t) = -\vec{\Lambda}[\vec{f}](s)- \vec{K}[\vec{f}](s),
\end{align}
where the local, $\vec{\Lambda}$, and the non-local operators, $\vec{K}$, are  respectively given by
\begin{align}\label{operators1}
\vec{\Lambda}[\vec{f}](s) &= [-c(\vec{I} + \vec{\hat{s}}\vec{\hat{s}}) + 2(\vec{I} - \vec{\hat{s}}\vec{\hat{s}})] \cdot \vec{f}(s), \\
\label{operators2}
\vec{K}[\vec{f}](s) &= \int_{0}^{L} \left(\vec{G}(s,s^\prime) \cdot \vec{f}(s^\prime) - \dfrac{\vec{I} + \vec{\hat{s}}\vec{\hat{s}}}{|s-s^\prime|}\cdot \vec{f}(s)\right) \,\mathrm{d}s^\prime,
\end{align}
where $\vec{G}$ is the Stokeslet singularity defined as 
\begin{align}
G_{ij}(s,s^\prime) = \dfrac{\delta_{ij} + \hat{R}_i\hat{R}_j(s,s^\prime)}{R(s,s^\prime)},~ \vec{R} = \vec{x}(s) - \vec{x}(s^\prime),~\vec{\hat{s}}=\frac{\partial \vec{x}(s)}{\partial s},~\vec{I}_{ij}=\delta_{ij},
\end{align}
where $i,j=\{1,2,3\}$, $s$ is the parametric arc length of the object and the constant $c=\log(\epsilon^2 e)$. The parameter that defines slenderness is $\epsilon = r/L$, $r$ and $L$ being the cross-sectional radius and total arc length of the slender object. 
  
We solve Eq.~\eqref{sbt}   numerically in order to obtain the force distribution on a helix that undergoes rigid-body rotation. The helix is divided into small straight segments over which the force and velocity are assumed to be constant. This gives rise to a linear system of the form $u_i = M_{ij} f_j$ that can be solved for the velocity if the force distribution is known, or inverted to find the force distribution if the velocity is known (which is the case here since the rotation is being prescribed).

\begin{figure}[t]
	\includegraphics[width=0.7\textwidth]{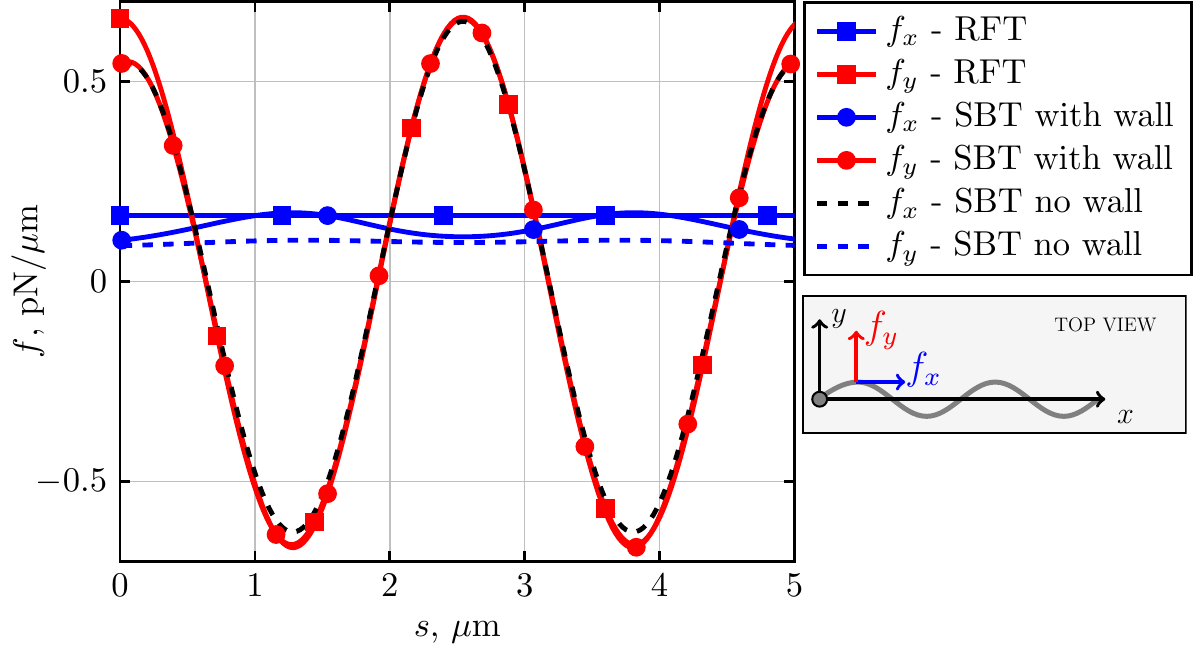}
	\caption {Comparison between the analytical (RFT) and numerical (SBT) results for the components of the hydrodynamic force density 
		$\vec{f}$ as a function of arc length, $s$, for {\it E.~coli}'s helical filament  rotating above the no-slip wall at   height $h=0.65~\mu$m. Geometrical values are taken from Ref.~\cite{gao2015using}.} 
	\label{fig:2}
\end{figure}

In order to account for the presence of wall we place the slender helix in a semi-infinite domain above a rigid boundary. The first term in the non-local operator $\vec{K}$, Eq.~\eqref{operators2}, contains the self-interactions of the slender filament. 
To include the presence of  the rigid wall at $z=0$, we place image singularities for a Stokeslet \citep{blake1974} on the opposite side of the wall for each element of the slender object so as to satisfy the no-slip boundary condition on the wall exactly. The operator $\vec{K}$ is then modified as, 
\begin{align}
\vec{K}[\vec{f}](s) &= \int_{0}^{L} \left([\vec{G}(s,s^\prime) + \vec{G}^{im}(s,s^\prime)] \cdot \vec{f}(s^\prime) - \dfrac{\vec{I} + \vec{\hat{s}}\vec{\hat{s}}}{|s-s^\prime|}\cdot \vec{f}(s)\right) \,\mathrm{d}s,
\end{align}
where $\vec{G}^{im}$ is the image of the Stokeslet,
\begin{align}
G_{ij}^{im} = -\dfrac{\delta_{ij} + \hat{R}^{im}_i\hat{R}^{im}_j}{R^{im}} + 2h \Delta_{jk}\dfrac{\partial}{\partial R^{im}_k}\left(\dfrac{h\hat{R}^{im}_i}{R^{im2}} - \dfrac{\delta_{i3} + \hat{R}^{im}_i\hat{R}^{im}_3 }{R^{im}}\right),
\end{align}
where $h=R_3$ is the distance from the wall, $\vec{R}^{im}(s,s^\prime) = \vec{x}(s) - \vec{x}^{im}(s^\prime)$, and $\vec{x}^{im} = [R_1,R_2,-R_3]$. The tensor $\Delta_{jk}$ takes the value of 1 when $j=k=1,2$ and -1 when $j=k=3$ and is 0 for every other combination. We can then solve Eq.~\eqref{sbt} numerically as above using   the modified operator $\vec{K}$. 
The obtained force distribution $\vec{f}$ is shown in  Fig.~\ref{fig:2}.
 We next integrate the force density on the fluid, $\vec{f}$,  along the arc length of the helix to obtain to the total force and torque along the helix axis and the resulting flow rate, 
\begin{align}
\vec{F}&=\int_0^{L}\vec{f}ds,\\
\vec{G}&=\int_0^{L}
R(\vec{e}_{\sigma}\times\vec{f})ds,\\
\vec{Q}&=\frac{1}{\pi\mu}\int_0^{L}\vec{f}_{\parallel}h(s)ds,
\end{align}
where $(\vec{e}_{\sigma},\vec{e}_{\theta},\vec{e}_{x})$ are the cylindrical coordinates with the flagellar axis along $x$ and $\vec{f}_{\parallel}$ is the  force density component parallel to the no-slip wall at $z=0$, i.e.~$\vec{f}_{\parallel} = \vec{f}-(\vec{f}\cdot \vec{e}_z)\vec{e}_z $.

\subsection{Comparing analytical model with computations}

We may compare the results from the analytical RFT model with the numerical SBT approach by using the set of parameters applicable to {\it E.~coli} bacteria~\cite{gao2015using}, namely $P=2~\mu$m, $R=0.25~\mu$m, $L_{axial}=4~\mu$m, $r=12$~nm, $f=165$~Hz, $h=0.65~\mu$m. Force distributions parallel (i.e.~along the $x$ direction) and perpendicular to the helical axis (i.e.~along the $y$ direction) are shown in Fig.~\ref{fig:2}. The main difference between the SBT with and without the hydrodynamic presence of the wall is that adding the wall slightly increases the force component $f_x$ and leads to small end effects. The analytical model (RFT) which is much  simpler to implement and   does not include the effect of the wall is in  good agreement with   SBT despite a small but systematic overestimation of the force $f_x$. Interestingly, the forces parallel to the no-slip wall, i.e.~$f_x,f_y$, are not very sensitive to the presence of the wall.
\subsection{Comparison between theory and experiments}
\begin{figure}[t]
\includegraphics[width=0.55\textwidth]{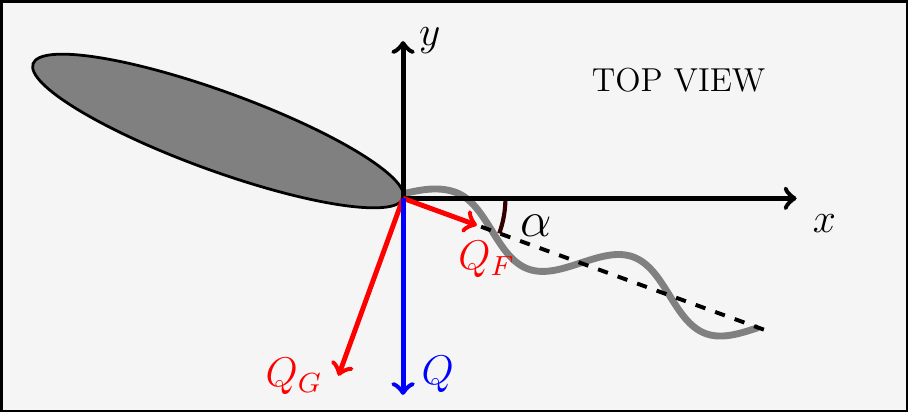}
\caption {Top view of a bacterium with a rotating flagellar filament. The angle between the axis of the filament and the normal to direction of the net flux $Q$ is denoted by $\alpha$. The  flow rate  due to the force has magnitude  $Q_F$ while that  due to the torque is denoted $Q_G$.}
\label{fig:10}
\end{figure}

Using the parameters for {\it E.~coli}~\cite{gao2015using}, we may compute the magnitude of the flow rate, $Q$,  and its direction, $\beta$, produced  by the rotating flagellar filament.  Experimentally, one  would like to pump the fluid along a specific direction at the maximum rate,  say the negative $y$ direction. If we define the angle $\alpha$ as the angle between the helix axis and the $x$ axis such that the flux vector is along negative $y$ direction (Fig.~\ref{fig:10}) and given the definition of $\beta$ shown in Fig.~\ref{fig:1} (top) then one should position the cell at the optimal angle $\alpha$ such that  $\alpha=\pi/2+\beta$.
We use our theory and computations to compute the value of this optimal angle, $\alpha$, and the resulting magnitude of the flow rate, $Q$.  We obtain results in excellent agreement with each other, namely
\begin{align}
\text{RFT}:~\alpha=52.5^{\circ},~Q=219~\mu\text{m}^3/\text{s},~F=0.838~\text{pN},
~G=835~\text{pN}\cdot\mu\text{m},\\
\text{SBT}:~\alpha=50.6^{\circ},~Q=182~\mu\text{m}^3/\text{s},~F=0.712~\text{pN},
~G=832~\text{pN}\cdot\mu\text{m},
\end{align}
\begin{figure}[t!]
\includegraphics[width=0.495\textwidth]{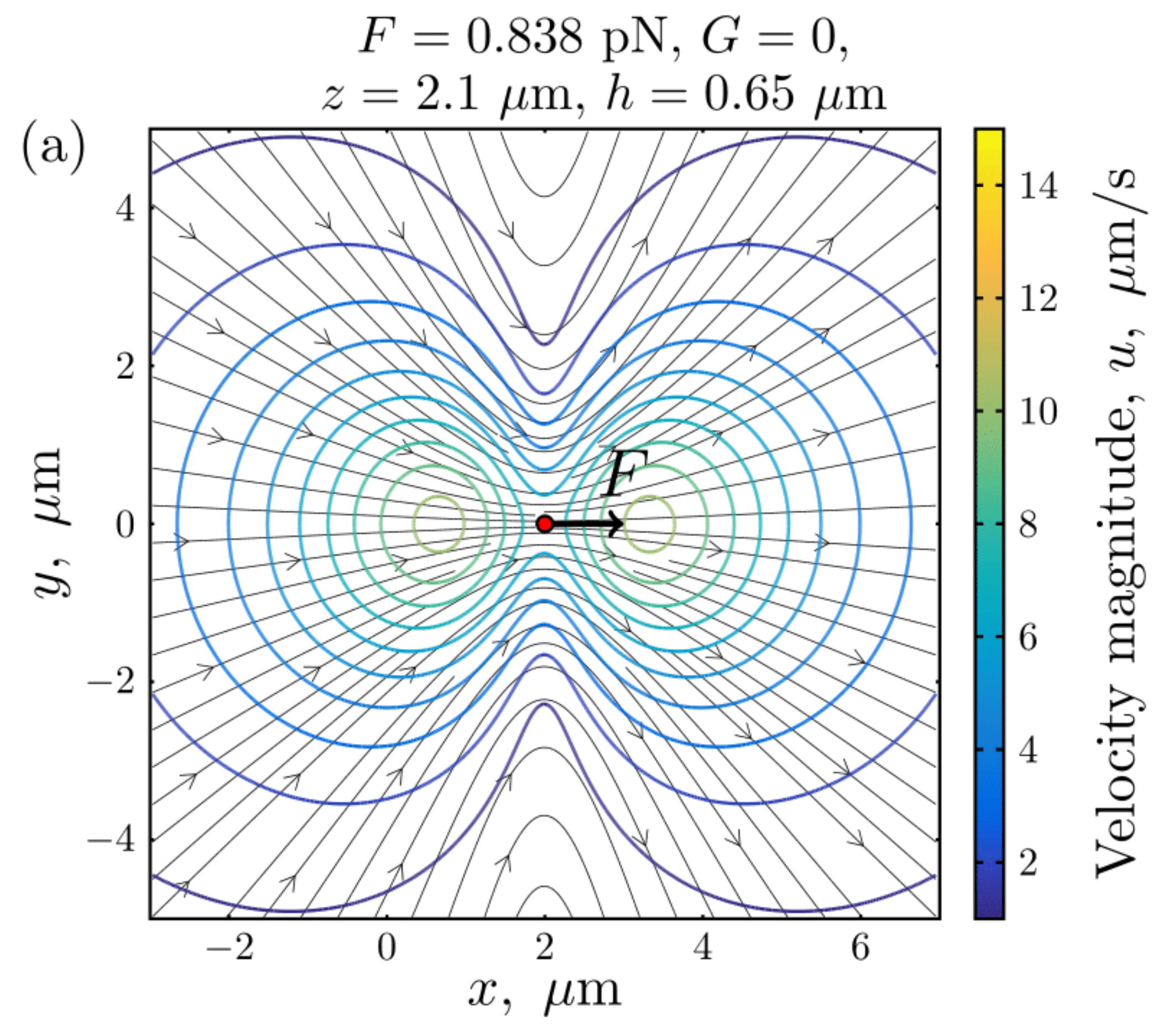}
\includegraphics[width=0.495\textwidth]{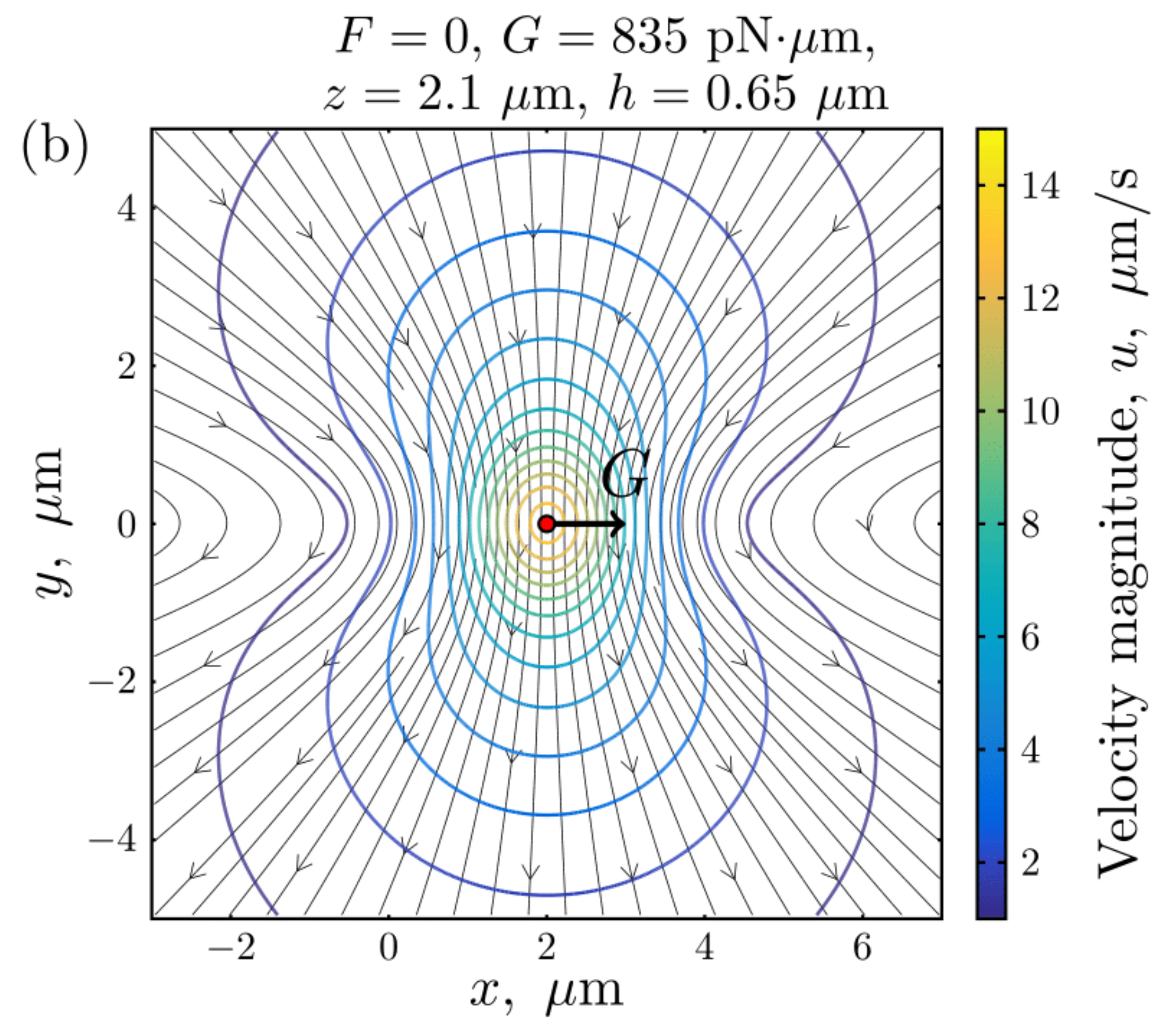}
\includegraphics[width=0.495\textwidth]{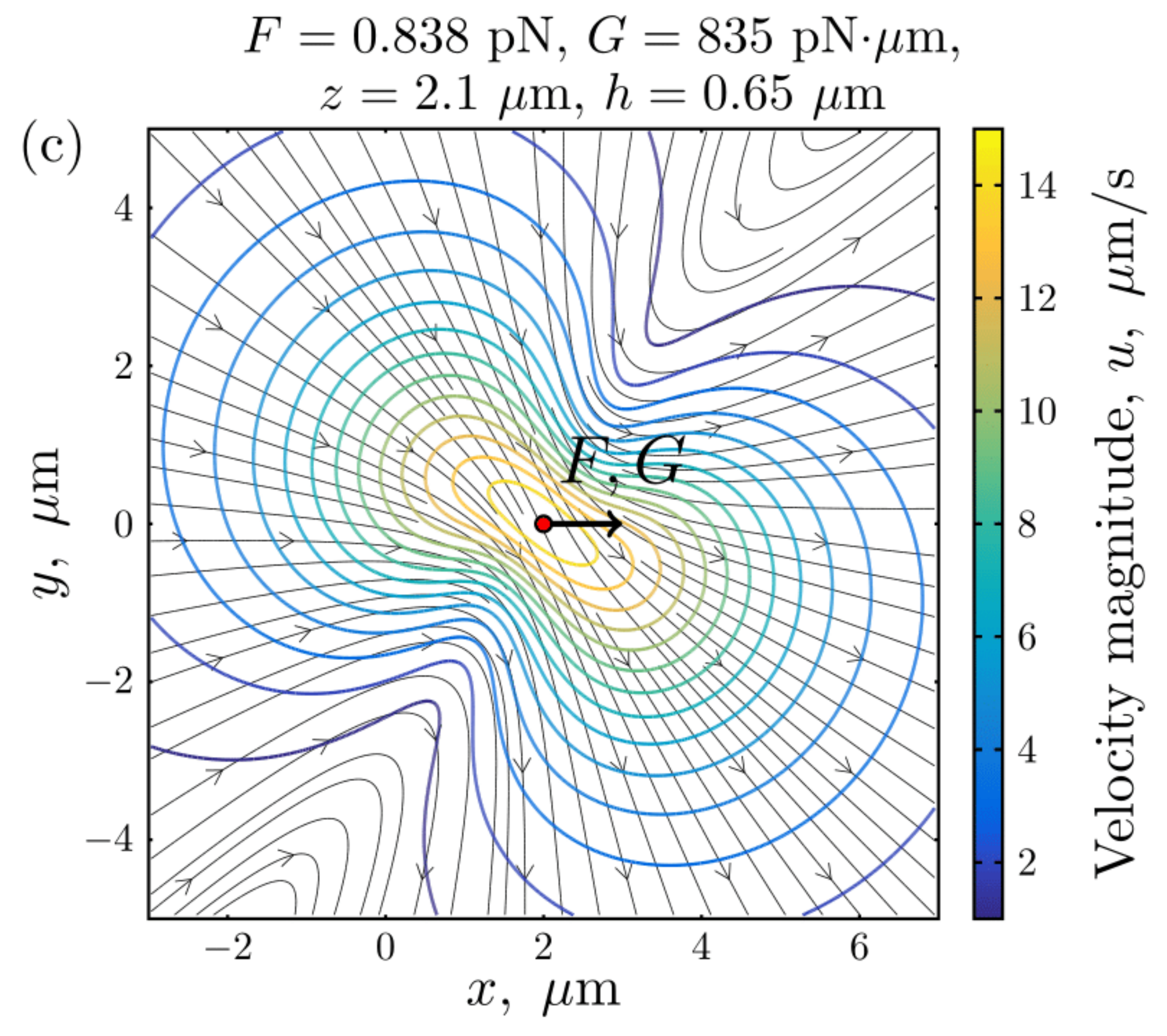}
\caption {Streamlines and contour lines of the two-dimensional velocity magnitude  for the helix modelled as a point force $F$ and a point torque $G$   at the height $h$ above the no-slip wall. (a) Fluid flow due to the point force only; (b) Flow due to  torque only; (c) Combined flow due to both singularities.}
\label{fig:5}
\end{figure}
In  Ref.~\cite{gao2015using}, Gao et.~al.~conducted experiments with \textit{E.~coli} bacteria and performed numerical simulations on the fluid flow due to trapped bacteria which were placed at angles $0^{\circ},~20^{\circ},~40^{\circ},~60^{\circ}$ with respect to the $x$ axis (using our notation). It was found that the among those values, the maximum flow rate was obtained 
for an angle of $40^{\circ}$. Our theoretical approach agrees with their results.   Furthermore we are able to predict that the configuration with $\alpha\approx50^{\circ}$ is the best one for the use of these cells to pumps fluids.
\subsection{Fluid flow visualisation}
In order to gain further understanding of the flow due to the superposition of the two flow singularities above the no-slip wall (Stokeslet and Rotlet)  we plot in Fig.~\ref{fig:5} the streamlines and contour lines of 
  the velocity magnitude, $u=\left(u_x^2+u_y^2\right)^{1/2}$,  on the plane at height $z=2.1~\mu$m above the wall (with geometrical parameters from 
  Ref.~\cite{gao2015using}). Specifically, we plot in Fig.~\ref{fig:5}(a) 
   the flow due to the point force only, in Fig.~\ref{fig:5}(b) the flow due to the point torque only and Fig.~\ref{fig:5}(c) shows the superposition of these two flows. As  expected the combination of pushing of the fluid along the helix and rotating the helix, we clearly observe a flow occurring  at an angle to the helix orientation. 

\begin{figure}[t!]
\includegraphics[height=0.55\textwidth]{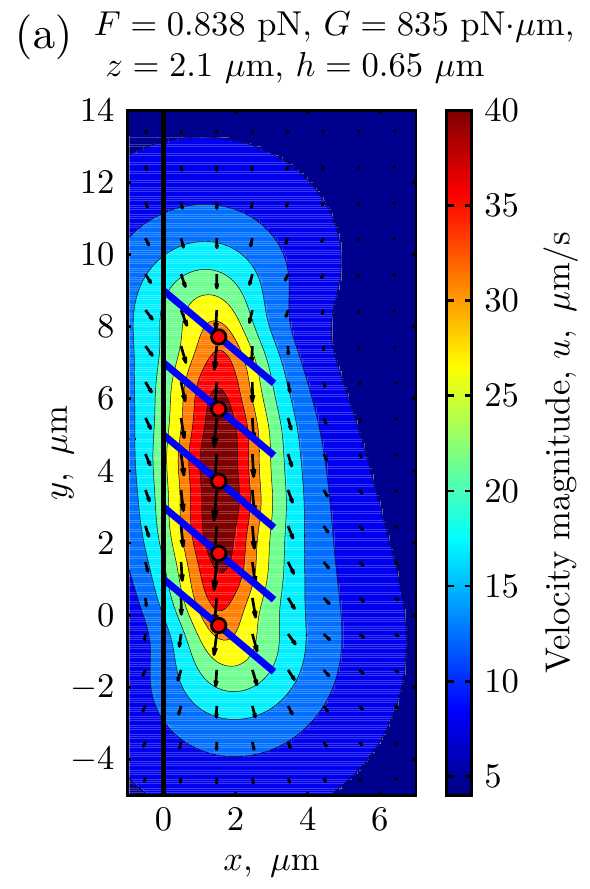}
\includegraphics[height=0.55\textwidth]{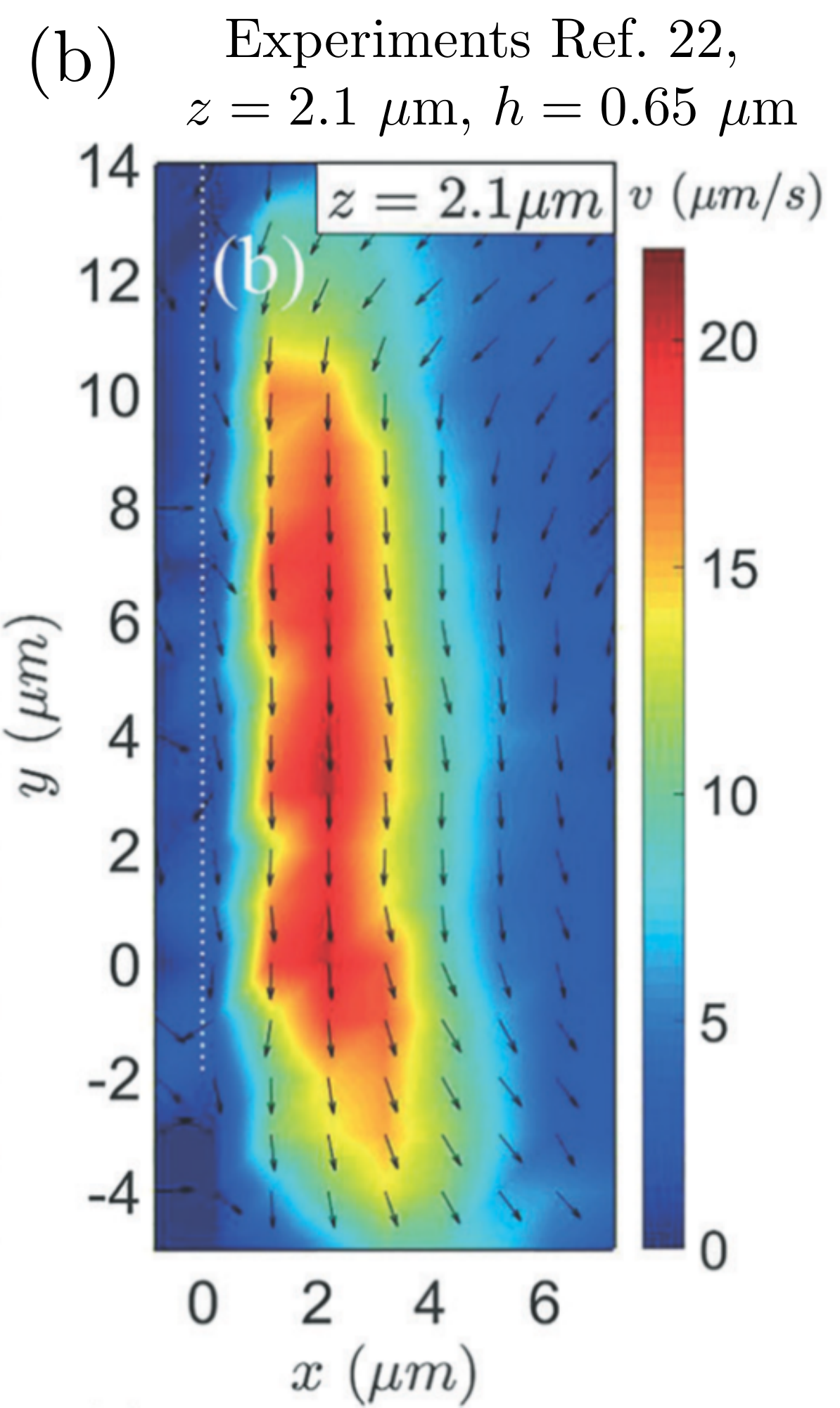}
\caption {Velocity vectors and 2D velocity magnitude contours for five helices tilted at angle $\alpha=40^{\circ}$, at height $h=0.65~\mu$m above the wall. The flow is shown above the helices in the plane  $z=2.1~\mu$m. The helical axes are denoted by thick blue lines with the red dots in the middle. (a): Flow obtained using the analytical model (RFT) as the superposition of flows due to individual helices; (b): Flow measured  experimentally. Reproduced from Ref.~\cite{gao2015using} by permission of The Royal Society of Chemistry.} 
\label{fig:6}
\end{figure}

A comparison between the flow field predicted by the analytical RFT model in the case when we pick   $\alpha=40^{\circ}$ and the experimental measurements of Ref.~\cite{gao2015using} in the same configuration is shown in Fig.~\ref{fig:6}. The flow field predicted by the theory is qualitatively similar to the experimentally obtained one and in both cases we can clearly see the localised flow occurring mostly along the $y$ axis. Note that the  experimental flow is about half as strong as the theoretically predicted one which is due to the  fluid  being stopped by the presence of the boxes which trap bacteria and which are absent in our theory (they essentially provide a vertical no-slip wall at $x=0~\mu$m).
  
\subsection{Dependence on the distance to the wall}
\begin{figure}[t]
\includegraphics[width=0.495\textwidth]{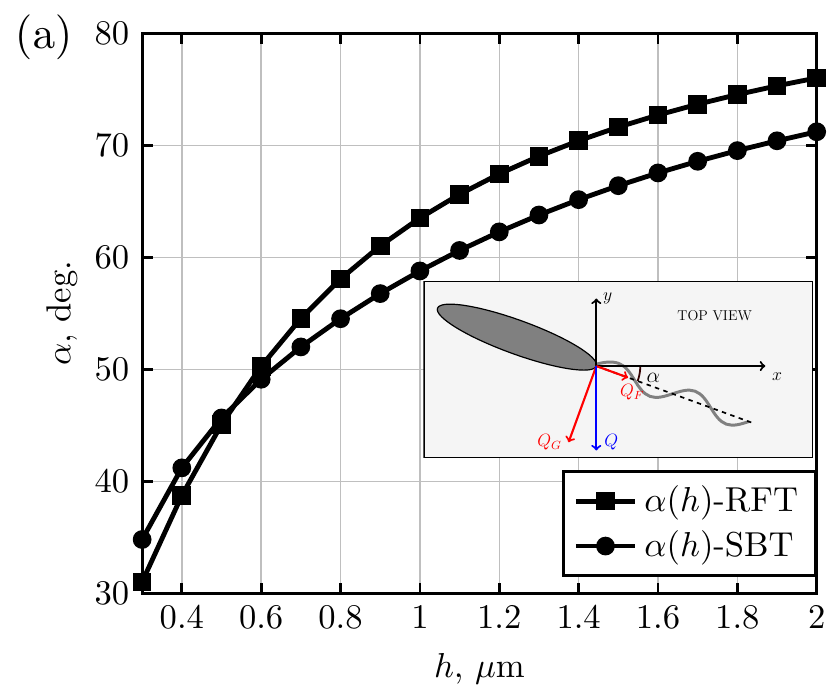}
\includegraphics[width=0.495\textwidth]{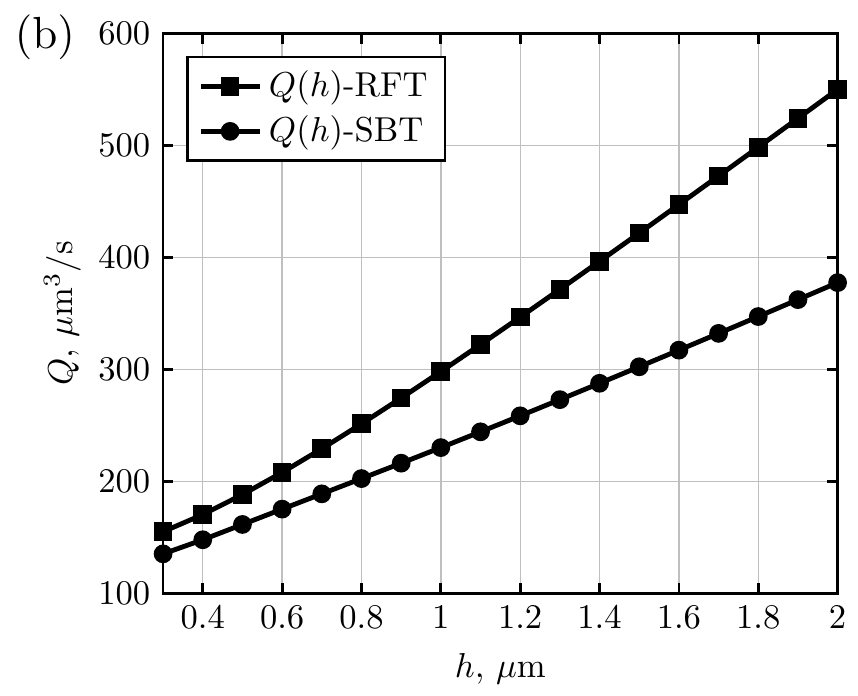}
\caption {Comparison between the analytical model (RFT) and the numerical approach (SBT) for (a) the optimal angle $\alpha$ and (b) the flux $Q$ as a function of height $h$ for the helix rotating above the no-slip wall.} 
\label{fig:3}
\end{figure}

So far we have assumed that the distance between the cells and the surface was a fixed height, $h$. If the helical filament was placed  closer or further from the bottom surface, how would this impact the value of the  optimal angle $\alpha$? Our theoretical and computational models (RFT and SBT) allow us  to investigate how the flux magnitude, $Q$, and the optimal angle, $\alpha$, depend on the value of $h$, with results shown in  Fig.~\ref{fig:3}.
We obtain that the analytical and numerical results are in good agreement. Furthermore, the analytical model predicts the optimal angle using to Eq.~\eqref{eq:beta} is given by
\begin{align}
\alpha&=\arctan{\left(\frac{ 2 F h }{G}\right)}=\arctan{\left(\frac{ 2 (1-\rho) \sin{\Psi} \cos{\Psi}h }{(\cos^2{\Psi}+\rho \sin^2{\Psi)}R}\right)},
\end{align}
which quantifies   the relative strength between  the flux generated by the helix torque, $G$, and that  due to the force above the wall, $2Fh$. We see that the flow rate is systematically  overestimated by the analytical model compared to the simulations though the  optimal angle $\alpha$ is almost identical for both models. This overestimation using the RFT is due to the overestimation of the total force $\vec{F}$ which becomes more important for larger $h$.
\section{Optimal helical form}

While the experiments in Ref.~\cite{gao2015using} exploited the flagella of  \textit{E.~coli} in order to generate  flows, the recent development of artificial bacteria flagella~\cite{zhang2010artificial,tottori2012magnetic}   suggests that helical shapes different from the biological ones could also be   used in order to induce pumping. In this section we investigate theoretically if a better (artificial) helical filament could be used to drive the flow.

We pose the optimisation problem in the following manner. We wish to  maximise the magnitude of the flow rate, $Q$, given a constrain on the rate of energy dissipation in the fluid, $D$ (which is equal to the rate of working on the rotating helix against the fluid). Using the analytical RFT approximation, the magnitude of the flow rate is given by 
\begin{align}
Q^2&=\left(\frac{F h}{\pi\mu}\right)^2+\left(\frac{G}{2\pi\mu}\right)^{2},\\
 \nonumber
&=\left(\frac{\xi_{\perp} R \omega (1-\rho) \sin{\Psi} \cos{\Psi} L h}{\pi\mu}\right)^2+\left(\frac{\xi_{\perp} R^2 \omega (\cos^2{\Psi}+\rho \sin^2{\Psi)} L}{2\pi\mu}\right)^{2},
\end{align}
while the rate of viscous dissipation in the fluid  due to the  rotating but not translating helix using Eq.~\eqref{Gx} is
\begin{align}
D=G\omega=\xi_{\perp} R^2 \omega^2 (\cos^2{\Psi}+\rho \sin^2{\Psi)} L.
\end{align}

Let us assume that we wish to keep the dissipation constant, i.e.~$D=D_0$, and maximise the magnitude of $Q$. Substituting $L$ in terms of $D_0$ into the  expression for the flux one gets the flow rate now given by
\begin{align}
Q^2=\left[1+\left(\frac{2h(1-\rho)\sin{\Psi}\cos{\Psi}}{R(\cos^2{\Psi}+\rho\sin^2{\Psi})}\right)^2\right]\left(\frac{D_0}{2\pi\mu\omega}\right)^2.
\end{align}
We see that the flux $Q$ is inversely proportional to the angular velocity $\omega$, which means that for a given dissipation, $D_0=G\omega$, smaller a angular velocity $\omega$ will lead to a bigger torque $G$, i.e.~a larger flow rate. This  can be achieved, for example, by having a longer helix and thus a larger value of $L$.  

\begin{figure}[t!]
	\includegraphics[height=0.44\textwidth]{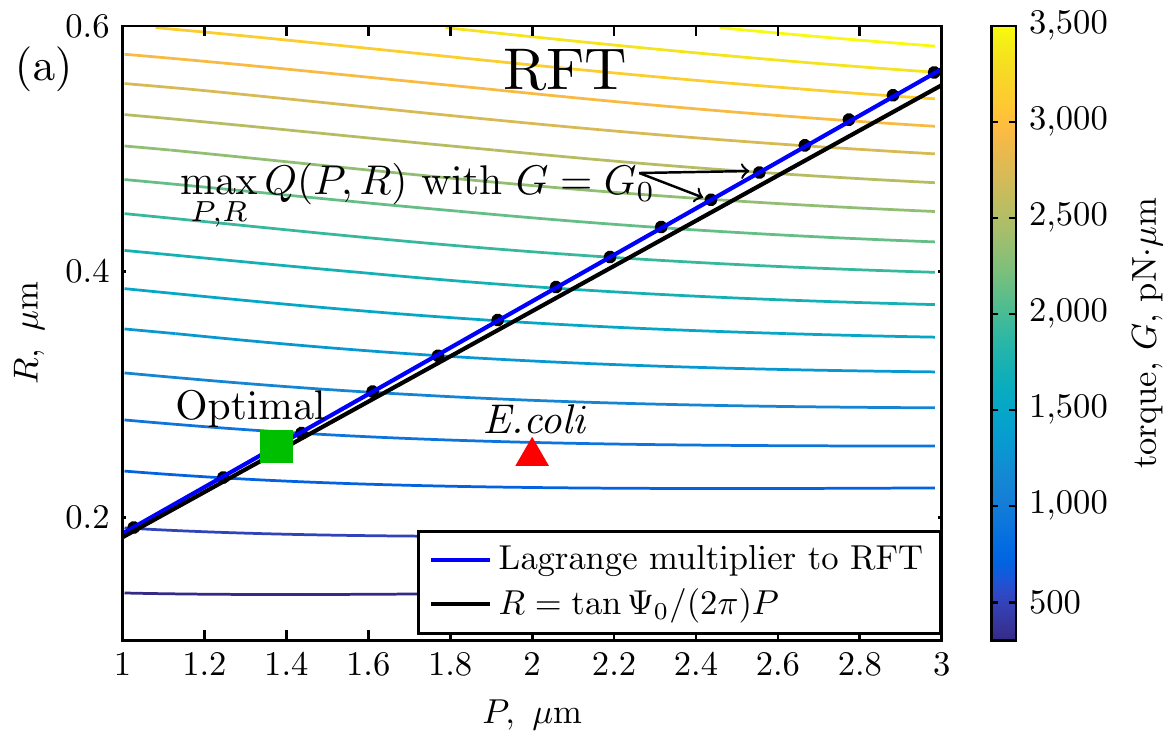}
	\includegraphics[height=0.44\textwidth]{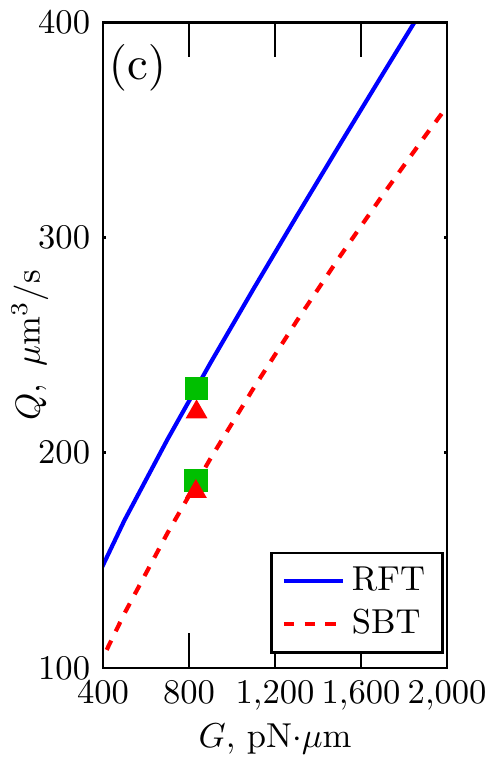}
	\includegraphics[height=0.44\textwidth]{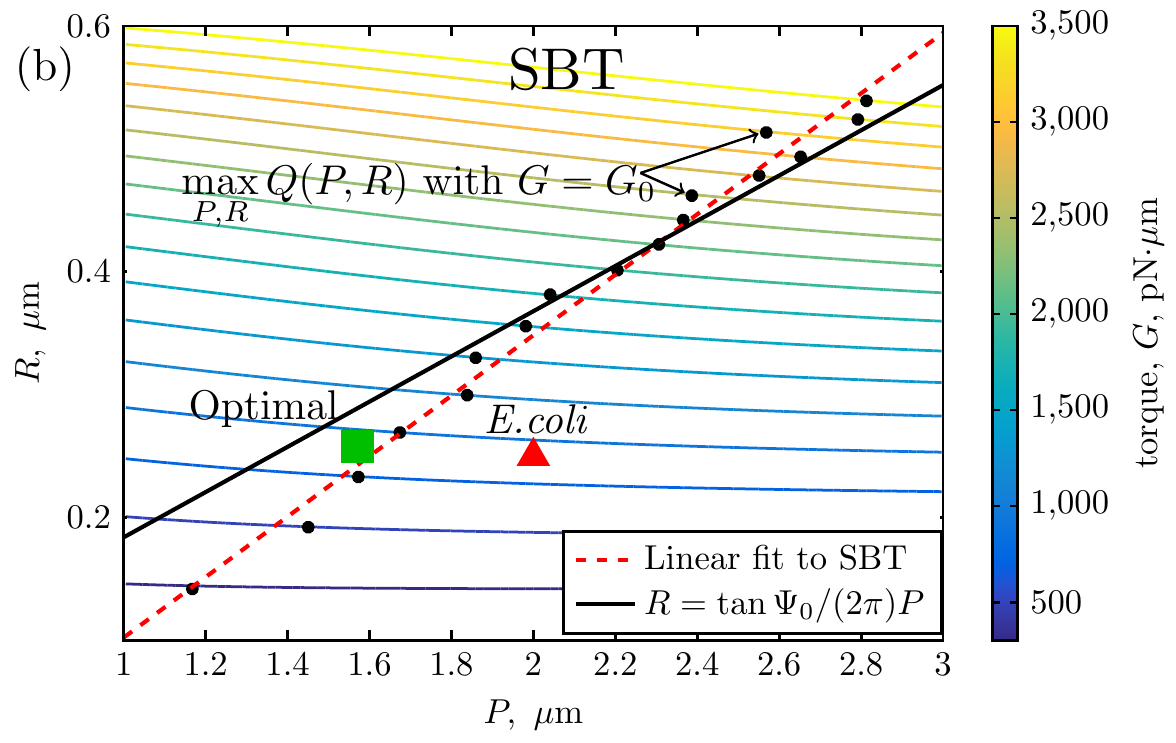}
	\includegraphics[height=0.43\textwidth]{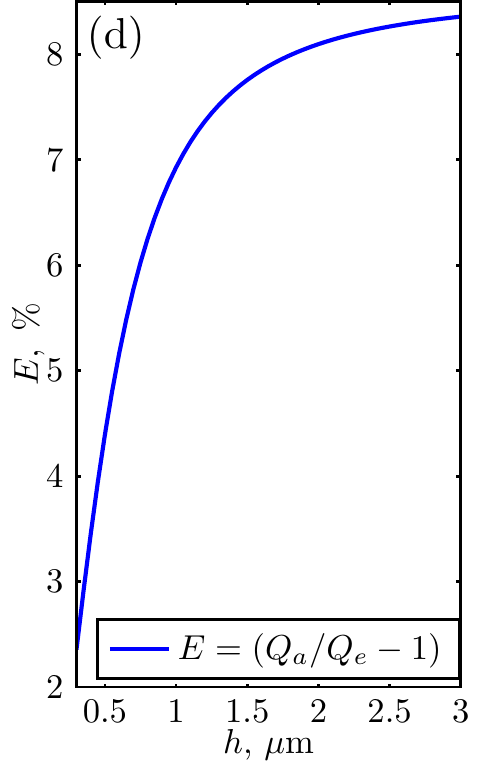}
	\caption {(a) and (b): Contours values of the torque,  $G$,  as a function of the helix radius, $R$, and pitch, $P$. Torques computed using the analytical (RFT) model (a) and the numerical (SBT) model (b). The scatter plots show the discrete values of $R,P$ that maximises  $Q$ given the torque values $G_0=500,700,...3300$~pN$\cdot\mu$m.  (c) Optimal flux, $Q$, as a function of the applied torque, $G$. (d) Marginal increase in the flux, $E$, using the optimal helix versus the  \textit{E.~coli} helix as a function of distance to the wall, $h$. In (a), (b), (c) the red triangles denote the values of pitch, radius, torque and flux for the \textit{E.~coli} bacterium  while the green square shows the optimal radius, pitch and flux for  the same torque as  \textit{E.~coli}.}
	\label{fig:7}
\end{figure}

For a given total torque on the helix $G$, what is the shape of the  helix which produces the biggest flux $Q$? In order to find the answer,   we need to maximise the force $F$ for the fixed height $h$. We know that $F$ is linearly proportional to the angular speed $\omega$ and the arc length $L$. Fixing the values of $G,\omega, h, L$ and $r$ sets the dissipation $D_0=G\omega$, and we aim to determine   the helix radius, $R$, and its pitch, $P$, so that the flow rate is maximal.
This minimisation problem may be formulated using the Lagrangian $M$ and the Lagrange multiplier $k$ given by
\begin{align}
M(P,R;k)&=Q^2(P,R)+k[G(P,R)-G_0],\\
Q^2&=\left(\frac{F h}{\pi\mu}\right)^2+\left(\frac{G}{2\pi\mu}\right)^{2},\\
 F&=\xi_{\perp} R \omega (1-\rho) \sin{\Psi} \cos{\Psi} L, \\
 G&=\xi_{\perp} R^2 \omega (\cos^2{\Psi}+\rho \sin^2{\Psi)} L,\\
 \xi_{\parallel} &=\frac{2\pi\mu}{\ln{(0.18 P\sec{\Psi}/r)}},
 \xi_{\perp} =\frac{4\pi\mu}{1/2+\ln{(0.18 P\sec{\Psi}/r)}},~\rho=\frac{\xi_{\parallel}}{\xi_{\perp}},
\end{align}
We numerically solve the resulting Euler-Lagrange equations based on the RFT model
\begin{align}
\frac{\partial}{\partial P}M(P,R;k)=0,~\frac{\partial}{\partial R}M(P,R;k)=0,~\frac{\partial}{\partial k}M(P,R;k)=0,
\end{align}
in order to determine the extrema for the flux. The solution is shown in Fig.~\ref{fig:7}(a) as a scatter plot for  torques in the range~$G_0=500,700,...3300$~pN$\cdot\mu$m. 

The extrema points $(P,R)$ all lie on the fitted line $R=0.189P-0.002~\mu$m which can be rationalised as follows. Ignoring the weak logarithmic dependence of $\xi_\parallel, \xi_\perp$  on $R, P$ we can in fact find  analytically the optimal values of $R, P$ which maximise the flux for a given torque namely
\begin{align}
\frac{2\pi R}{P}=\tan{\Psi_0}=\rho^{-1/4}.
\end{align}
This analytical solution is shown as a black line in Fig.~\ref{fig:7}(a) and is very close to the blue line (numerical calculations).  
In this figure we also use the red triangle to denote the radius and pitch for the flagellar filament of  \textit{E.~coli} while the green square denotes the optimal radius and the pitch for pumping such that the torque is the same as that generated the \textit{E.~coli} bacterium. In the case of \textit{E.~coli}, it was reported that the torque applied by the rotary motor  remains
approximately constant up to a relatively high rotation frequency ($170$~Hz at  $23^{\circ}$C)~\cite{chen2000torque,nord2017speed}, and we use this biological value of 830 pN$\cdot\mu$m for $G$. 
We may then calculate the radius, pitch, flux and the optimal angle in terms of the fixed quantities for the optimal shape, i.e. when $\tan{\Psi}=\rho^{-1/4}$ and obtain
\begin{align}
R&=\left(\frac{G_0}{\xi_{\perp}\omega L} \right)^{1/2}\cdot\rho^{-1/4},~P=2\pi\rho^{1/4} R=2\pi\left(\frac{G_0}{\xi_{\perp}\omega L} \right)^{1/2},\\
Q^2&=\left(\frac{\xi_{\perp}R L\omega(\rho^{1/4}-\rho^{3/4}) h}{\pi\mu}\right)^2+\left(\frac{G_0}{2\pi\mu}\right)^{2},\\
Q^2&=\left(\frac{\xi_{\perp}^{1/2}L^{1/2}\omega^{1/2}G_0^{1/2} (1-\rho^{1/2})h}{\pi\mu}\right)^2+\left(\frac{G_0}{2\pi\mu}\right)^{2},\\
\tan{\alpha}&=2\xi_{\perp}^{1/2}L^{1/2}\omega^{1/2}G_0^{-1/2} h(1-\rho^{1/2}).
\end{align}
Surprisingly, the optimal pitch $P$ is independent of the drag ratio $\rho$ but is only a function of the applied torque $G_0$, the helix length $L$, the rotation frequency $\omega$ and the drag coeficient $\xi_{\perp}$. 

Instead of the analytical modelling approach, we may instead compute  the flux $Q$ as a function of the parameters of the helix, $R, P$, using the numerical model (SBT) with   fixed helix height above the wall $h=0.65~\mu$m. These results are shown in Fig.~\ref{fig:7}(b). The contours of the torque $G$ as the function of the helix radius $R$ and the pitch $P$ match well the ones calculated using the analytical model (RFT). In both cases, the contours are almost horizontal when the radius $R$ is small, e.g.~about $0.25~\mu$m in the case of \textit{E.~coli}. This can be justified by examining  the form of torque,
\begin{align}
G&=\xi_{\perp} R^2 \omega (\cos^2{\Psi}+\rho \sin^2{\Psi)} L=\xi_{\perp} R^2 \omega L \left(\frac{P^2+4\pi^2 \rho R^2}{P^2+4\pi^2 R^2}\right).
\end{align}
The torque depends weekly on the pitch $P$, especially when the helix radius $R$ is small. This means that the contours of $G$ are approximately $G(R,P)\sim R^2 =\text{const}$, i.e.~horizontal contour lines. 
The optimal value of the flux, $Q$, is plotted in Fig.~\ref{fig:7}(c)
as a function of the applied torque,  $G$. We see that the analytical  model systematically overestimates the flux compared with the computational approach. Remarkably, the difference between the flow rate produced by the \textit{E.~coli} bacterium and the optimal one for the same applied torque is only $3\%$, revealing  that the bacterial flagellar filaments are almost at the optimal shape to pump the fluid over the surface. 
How does this result depend on the distance to the surface?
We may compute  the ratio between the flux $Q_a$ due to the optimal artificial helix (force $F_a$, torque $G_0$) versus the flux $Q_e$ due to the \textit{E.~coli} shaped helix (force $F_e$, torque $G_0$).
\begin{align}
E &= \frac{Q_a}{Q_e} -1 = \left(\frac{G_0^2/(2\pi\mu)^2+F_a^2 h^2/(\pi\mu)^2}{G_0^2/(2\pi\mu)^2+F_e^2 h^2/(\pi\mu)^2}\right)^{1/2}-1,\\
E &=\left(1+\frac{F_a^2 h^2-F_e^2 h^2}{G_0^2/4+F_e^2 h^2}\right)^{1/2}-1=\left(1+\frac{(F_a/F_e)^2-1}{1+G_0^2/(4F_e^2 h^2)}\right)^{1/2}-1.
\end{align}

The applied torque calculated analytically  is $G_0=835~$pN$\cdot\mu$m. The biological force is $F_e =0.838~$pN  while the optimal force is $F_a =0.910~$pN. We see in Fig.~\ref{fig:7}(d) that the percentage increase in flow rate between the biological value and the optimal value in always in the range  $3-8~\%$ and increases with the  height above the wall. Bacterial filaments thus seem to be nearly optimal for pumping fluid very close to a surface, a result which  could be important in the context of  efficient expansion of bacterial swarms~\cite{wu2011microbubbles}.

\section{Conclusion}
Our simple analytical (RFT) and computational (SBT) models  are able to capture the leading-order physics of the fluid flux produced by a rotating helical bundle of flagellar filaments near a no-slip wall. Fluid is being pumped along the flagellar axis because of the net force acting on the fluid and  in the direction perpendicular  to the flagellar axis because of the rotation in the presence of the wall. The force produces a flow rate $Q_F=Fh/(\pi \mu)$ which depends on the distance to the wall $h$, whereas the torque produces the flux $Q_G=G/(2\pi\mu$). The combined effect of force and torque implies that the net flow rate is directed at an angle  to the flagellum, whose value depends on the helix parameters and its distance to the wall. While our model is limited to the cases where the boundary affecting the fluid flow is a flat no-slip wall,  more advanced numerical techniques would  be needed to tackle  complex boundaries such as channels, corners, irregular surfaces. Importantly, our theoretical predictions agree with the numerical  and experimental results in Ref.~\cite{gao2015using}, and thus provide fundamental understanding of the relationship between orientation and flow pumping, setting the stage for the future development of  efficient micropumps not only using bacteria but also other actuation methods (e.g.~magnetic).

We next investigate the optimal shape of the helix in terms of its radius $R$ and pitch $P$ assuming a  fixed applied torque $G$, total helical length $L$, angular speed $\omega$, height above the wall $h$ and  filament radius $r$. 
We find that  nearly-optimal pumping is achieved when the helix pitch angle is fixed and equal to $\tan{\Psi}=2\pi R/P=\rho^{-1/4}$, where $\rho$ is the ratio between the drag coefficients. With a drag ratio $\rho=1/2,$ one finds that $\Psi$ is close to $50^{\circ}$. While the pitch angle for normal flagellar filaments is close to $40^{\circ}$ and is  optimised for propulsion~\cite{spagnolie2011comparative}, the  theoretically-optimal pumping helix  performs only $3~\%$ to $8~\%$ better than $\textit{E.~coli}$ depending on the height above the wall. This is   a very small difference indicating that  bacterial flagella are very efficient micropumps, a result which could be relevant to the expansion of bacterial swarms~\cite{kearns2010field} or the future development of  biological micropumps to be used for   automation in chemistry and biology~\cite{squires2005microfluidics}.
  
\section*{Acknowledgements}

This work was funded in part by an ERC Consolidator grant from the European Union (EL).
\appendix
\section{Point force above an infinite wall at $z=0$}
\label{appendix:force}
Conside  a Stokeslet placed at $\vec{y_0}=(0,0,h)$. Define $\vec{r}=\vec{x}-\vec{y_0}=(x,y,z-h)$ and $\vec{R}=(x,y,z+h)$. 
The solution for a Stokeslet in the vicinity of a stationary plane boundary is~\cite{blake1974,blake1974fundamental}
\begin{align}
 u_i &= \frac{F_j}{8 \pi \mu}\left[\left(\frac{\delta_{ij}}{r}+\frac{r_i r_j}{r^3}\right)- \left(\frac{\delta_{ij}}{R}+\frac{R_i R_j}{R^3}\right)\right]+
 \\ & \nonumber+\frac{F_j}{8 \pi \mu}\left [2h(\delta_{j \alpha}\delta_{\alpha k}-\delta_{j 3}\delta_{3 k}) \frac{\partial}{\partial R_k}\left\{\frac{h R_i}{R^3}-\left(\frac{\delta_{i3}}{R}+\frac{R_i R_3}{R^3}\right) \right\}\right].
\end{align}
We are interested in the Stokeslet   parallel to the wall, therefore without loss of generality we choose $\vec{F}=(F_1,0,0)$. Let us calculate the flux in the plane perpendicular to the Stokeslet
\begin{align}
 Q_1&= \int_0^{\infty}\int_{-\infty}^{\infty}u_1 dy\,dz.
\end{align} 
Expand the flow to the leading order in $h$ and integrate to get
\begin{align}
 u_1 =\frac{F_1}{8 \pi \mu}\left[\frac{12x^2 zh}{(x^2+y^2+z^2)^{5/2}}+\mathcal{O}(h^2)\right],~Q_1=\frac{F_1h}{\pi \mu}\cdot
\end{align}
Therefore, there is a finite flow rate produced in the direction of the Stokeslet  given by 
\begin{align}
\vec{Q}=\frac{\vec{F}_{\parallel}h}{\pi \mu},
\end{align}
where $\vec{F}_{\parallel}=(F_1,F_2,0)$. Note that the flux is zero if the force is perpendicular to the wall, as expected by symmetry.

\section{Point torque above an infinite wall at $z=0$}
\label{appendix:torque}
Consider now  a Rotlet placed at $\vec{y_0}=(0,0,h)$. Define $\vec{r}=\vec{x}-\vec{y_0}=(x,y,z-h)$ and $\vec{R}=(x,y,z+h)$. 
The solution for the Rotlet in the vicinity of a stationary plane boundary is~\cite{blake1974,blake1974fundamental}
\begin{align}
 u_i = \frac{G_j\epsilon_{ijk}}{8 \pi \mu}\left[\frac{r_{k}}{r^3}-\frac{R_{k}}{R^3} \right]+\frac{G_j\epsilon_{kj3}}{8 \pi \mu}\left [2h\left(\frac{\delta_{ik}}{R^3}-\frac{3R_i R_k}{R^5}\right)+\frac{6 R_i R_k R_3}{R^5}\right].
\end{align}
Take $\vec{G}=(0,G_2,0)$ then the flux produced by this Rotlet is
\begin{align}
Q_1&= \int_0^{\infty}\int_{-\infty}^{\infty}u_1 dy\,dz,~Q_2=0,~Q_3=0.
\end{align} 
Expanding the flow to the order in $h$ and integrate we obtain
\begin{align}
 u_1=\frac{G_2}{8 \pi \mu}\left[\frac{6x^2 z}{(x^2+y^2+z^2)^{5/2}}+\mathcal{O}(h)\right],~Q_1=\frac{G_2}{2\pi \mu}\cdot
\end{align}
So in general the flux given by
\begin{align}
\vec{Q}=\frac{\vec{G}_{\parallel}\times\vec{e}_z}{2\pi \mu},
\end{align}
where $\vec{G}_{\parallel}=(G_1,G_2,0)$ and here again the flux is zero if the torque is perpendicular to the wall.


\bibliographystyle{unsrt}
\bibliography{justas_thesis}

\end{document}